\documentclass[aps,twocolumn,prd,nofootinbib,showpacs]{revtex4-1}
\pdfoutput=1 % if your are submitting a pdflatex 
\usepackage{amsmath}
\usepackage{amsfonts}
\usepackage{bm}
\usepackage{graphicx}
\usepackage{hyperref}
\usepackage[normalem]{ulem}

\mathchardef\mhyphen="2D
\usepackage[T1]{fontenc} % if needed

\newcommand{\ie}{{i.e.~}}

\newcommand{\eg}{e.g.~}

\newcommand{\apriori}{{a priori~}}

\let\oldsqrt\sqrt
\def\sqrt{\mathpalette\DHLhksqrt}
\def\DHLhksqrt#1#2{%
\setbox0=\hbox{$#1\oldsqrt{#2\,}$}\dimen0=\ht0
\advance\dimen0-0.2\ht0
\setbox2=\hbox{\vrule height\ht0 depth -\dimen0}%
{\box0\lower0.4pt\box2}}

\newcommand{\order}[1]{\mathcal{O}\!\left(#1\right)}

\newcommand{\dd}{\mathrm{d}}
\newcommand{\ee}{e}

\newcommand{\calP}{\mathcal{P}}

\newcommand{\beq}{\begin{equation}}
\newcommand{\eeq}{\end{equation}}
\newcommand{\bea}{\begin{equation}\begin{aligned}}
\newcommand{\eea}{\end{aligned}\end{equation}}

\newlength{\wsingfig}
\newlength{\wdblefig}
\newlength{\wquadfig}
\newlength{\wtriplefig}
\newcommand{\Eq}[1]{Eq.~(\ref{#1})}
\newcommand{\Eqs}[1]{Eqs.~(\ref{#1})}
\newcommand{\Fig}[1]{Fig.~{\ref{#1}}}

\newcommand{\Refa}[1]{Ref.~{\cite{#1}}}

\newcommand{\Sec}[1]{Sec.~\ref{#1}}

\hypersetup{
    unicode=false,          % non-Latin characters in Acrobats bookmarks
    pdftoolbar=true,        % show Acrobats toolbar?
    pdfmenubar=true,        % show Acrobats menu?
    pdffitwindow=false,     % window fit to page when opened
    pdfstartview={FitH},    % fits the width of the page to the window
    pdfnewwindow=true,      % links in new window
    colorlinks=true,       % false: boxed links; true: colored links
    linkcolor=red,          % color of internal links
    citecolor=cyan,        % color of links to bibliography
    filecolor=magenta,      % color of file links
    urlcolor=blue,           % color of external links
    linktocpage=true
}

\begin{document} 
\flushbottom

\title{\boldmath Real-space entanglement of quantum fields}

\author{J\'er\^ome Martin${}^1$}
\email{jmartin@iap.fr}

\author{Vincent Vennin${}^{2,1}$}
\email{vincent.vennin@apc.in2p3.fr}

\affiliation{${}^1$Institut d'Astrophysique de Paris, UMR 7095-CNRS,
  Universit\'e Pierre et Marie Curie, 98 bis boulevard Arago, 75014 Paris, France\\${}^2$Laboratoire Astroparticule et Cosmologie,
  CNRS \& Universit\'e de Paris, 75013 Paris, France}

\date{\today}

\begin{abstract}
We introduce a new method permitting the analytical determination of entanglement entropy (and related quantities) between configurations of a quantum field, which is either free or in interaction with a classical source, at two distinct spatial locations. We show how such a setup can be described by a bipartite, continuous Gaussian system. This allows us to derive explicit and exact formulas for the entanglement entropy, the mutual information and the quantum discord, solely in terms of the Fourier-space power spectra of the field.
This contrasts with previous studies, which mostly rely on numerical considerations. As an illustration, we apply our formalism to massless fields in flat space, where exact expressions are derived that only involve the ratio between the size of the regions over which the field is coarse-grained, and the distance between these regions. In particular, we recover the well-known fact that mutual information decays as the fourth power of this ratio at large distances, as previously observed in numerical works. Our method leads to the first analytical derivation of this result, and to an exact formula that also applies to arbitrary distances. Finally, we determine the quantum discord and find that it identically vanishes (unless coarse-graining is performed over smeared spheres, in which case it obeys the same suppression at large distance as mutual information).
\end{abstract}

\maketitle

\section{Introduction}
\label{sec:intro}

Quantum systems differ from classical systems because of the way they are correlated. This simple remark has led to the development of several tools to quantify the amount and the nature of quantum correlations, both in discrete and continuous setups. Although such tools are often discussed in the context of systems containing one or a few degrees of freedom, their generalisation to setups containing infinitely many degrees of freedom allows one to study more realistic situations, and to discuss the case of quantum fields, which is required to tackle high-energy problems. 

An important result in this context is that the entanglement entropy of the reduced state of a quantum field inside a subregion often grows like the boundary area of that subregion~\cite{Srednicki:1993im, Calabrese:2004eu, Das:2005ah, Eisert:2008ur}, and not like its volume, which bears interesting similarity with the Bekenstein-Hawking intrinsic entropy of a black hole~\cite{Bekenstein:1973ur, Hawking:1974rv}. This has been shown to hold in a wide variety of setups, using different techniques.

In this work, we point out that, for quantum fields that are either free or interacting with a classical exterior source, the calculation of entanglement entropy and all derived quantities, such as mutual information or quantum discord, can be performed in a straightforward way, by making use of the Gaussian structure of the correlations. The idea is to describe the configurations of the field at two distinct spatial locations as a Gaussian bipartite system, for which tools have been developed that directly provide all relevant quantities~\cite{2014arXiv1401.4679A}. The correlation matrix of the bipartite system is simply given by the power spectra of the field, integrated against a window function that describes how the fields are locally coarse-grained. This leads us to explicit, analytical formulas solely in terms of the power spectra of the field.

This new method represents a significant improvement given that previous approaches were essentially based on numerical simulations. We then illustrate this formalism on a simple example. Even though we obtain new physical results (for instance, the calculation of quantum discord for Gaussian scalar fields is, to the best of our knowledge, new), this paper only represents a first step where we mainly aim at demonstrating that our setup is able to both confirm numerically-established results by an exact analytical calculation, and to study regimes (\eg at small distances) that are difficult to probe otherwise. Applications to new physical situations are considered elsewhere, for instance in~\Refa{Martin:2021qkg}.

The paper is organised as follows. In \Sec{sec:Gaussian}, we construct the bipartite systems associated to the configurations of quantum fields within two disjoint spheres. In \Sec{sec:GaussianStates}, we discuss how the von Neumann entropy, the mutual information and the quantum discord of such systems can be calculated explicitly from the knowledge of the two-point correlation functions of the field. We apply this generic framework to the case of a massless scalar field living in the Minkowski background in \Sec{sec:Minkowski}, where we show that the quantum discord identically vanishes. Finally, we present our conclusions in \Sec{sec:Conclusion}.

\section{Bipartite systems for two-point configurations of a quantum field}
\label{sec:Gaussian}

For simplicity, let us consider a quantum scalar field $\phi(\vec{x})$ living on a $D$-dimensional  spatial manifold, with conjugated momentum $\pi(\vec{x})$ (generalisation to other types of quantum fields can be carried out along similar lines). They satisfy the canonical commutation relation
\begin{align}
\label{eq:commutators:RealSpace}
\left[\phi(\vec{x}_1),\pi(\vec{x}_2)\right] = i \delta(\vec{x}_1-\vec{x}_2)\, .
\end{align}
The value of these fields averaged inside a sphere of radius $R$ is given by
\begin{align}
\label{eq:def:CoarseGrain}
\phi_R(\vec{x}) \equiv R^{-D}
\int\dd^D\vec{y} \,\phi (\vec{y})\, 
W\left(\frac{\left\vert \vec{y} - \vec{x}\right\vert}{R} \right) ,
\end{align}
and a similar expression for $\pi_R(\vec{x})$. In this formula, $W$ is a window function that singles out spatial points $\vec{y}$ distant from $\vec{x}$ by less than $R$. More precisely, we consider
\begin{align}
\label{eq:WindowFunction:Improved}
W(x)=\frac{1}{V_D {\cal F}(\delta)}
\begin{cases}
1 \quad \text{for} \quad x\le 1\, ,\\
\displaystyle
-\frac{1}{\delta}(x-1)+1 \quad \text{for}\quad 1<x\le 1+\delta\, , \\
0 \quad \text{for}\quad x>1+\delta \, ,
\end{cases}
\end{align} 
where $V_D=\pi^{D/2}/\Gamma(1+D/2)$ is the volume of the $D$-sphere of unit radius, $\Gamma(.)$ being the Euler function, and 
\begin{align}
{\cal F}(\delta)=\frac{(1+\delta)^{D+1}-1}{\delta (D+1)}
\end{align}
is set such that after coarse graining, a uniform field remains a uniform field of the same value. The window function is therefore a top-hat function within a sphere of radius $R$, to which a linear tail is added between $R$ and $R(1+\delta)$ that makes $W$ continuous. As we will show below, continuity is indeed required to properly account for mild UV divergences in some of the intermediate quantities we compute, although the limit $\delta\to 0$ will be taken in our final results. Let us also note that other smooth window functions could be used, but as we shall now see, in order for a bipartite system to be defined with canonical commutation relations, the window function needs to have a compact support and this makes the above choice natural. Other smooth, yet compact, window functions could obviously be considered, but this would not affect the limit $\delta\to 0$ where the field is coarse grained inside a sphere. 

\begin{figure}[t]
\begin{center}
\includegraphics[width=0.48\textwidth]{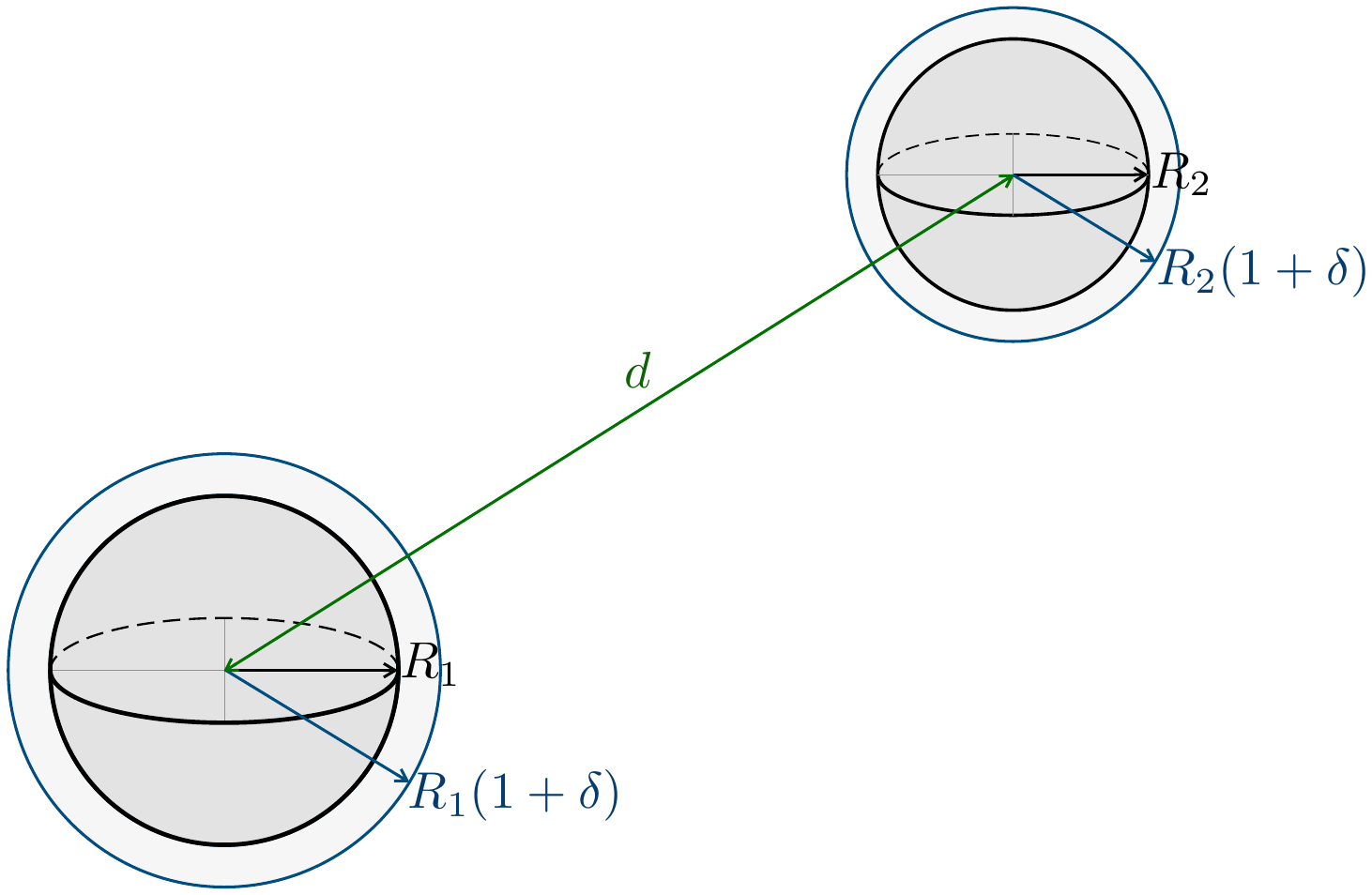}
\caption{Sketch of the setup studied in this work. A free quantum field is coarse grained within two non-overlapping spheres of radii $R_1$ and $R_2$, distant by $d$, and the correlations between the field configurations of these two spheres are studied. In practice, the coarse-graining window function is a top-hat function inside the spheres of radius $R_i$, with a linear tail between $R_i$ and $R_i(1+\delta)$. We also introduce the parameters $\kappa=\sqrt{R_1/R_2}$ and $\alpha=d/\sqrt{R_1 R_2}$ for convenience, where $\alpha>(\kappa+1/\kappa)(1+\delta)$ for the two regions to be disjoint.}
\label{fig:sketch}
\end{center}
\end{figure}
Let us now study the commutation relations between the field coarse-grained at $\vec{x}_1$ in a sphere of radius $R_1$, and the field coarse-grained at $\vec{x}_2$ in a sphere of radius $R_2$ (the situation is depicted in \Fig{fig:sketch}). Our goal is to see whether they can be cast in the same form as in \Eq{eq:commutators:RealSpace}. One obviously has $[\phi_{R_i}(\vec{x}_i),\phi_{R_j}(\vec{x}_j)]=[\pi_{R_i}(\vec{x}_i),\pi_{R_j}(\vec{x}_j)]=0$ for $i,j=1,2$. In order to have $[\phi_{R_1}(\vec{x}_1),\pi_{R_2}(\vec{x}_2)]=[\phi_{R_2}(\vec{x}_2),\pi_{R_1}(\vec{x}_1)]=0$, the support of the window functions centred at $\vec{x}_1$ and $\vec{x}_2$ must not intersect, so we restrict the analysis to pairs of points distant by 
\begin{align}
\label{eq:lowerbound:d}
d>(1+\delta)(R_1+R_2)\, .
\end{align} 
Making use of \Eq{eq:commutators:RealSpace}, one also finds $[\phi_{R_i}(\vec{x}_i),\pi_{R_i}(\vec{x_i})] = i D V_D R_i^{-D} \int_0^\infty u^{D-1} W^2(u) \dd u$. With \Eq{eq:WindowFunction:Improved}, the previous integral can be performed and this gives
\begin{align}
\label{eq:commutnoncan}
\left[\phi_{R_i}(\vec{x}_i),\pi_{R_j}(\vec{x}_j)\right] = i \frac{G(\delta)}{V_D R_i^D}  \delta_{ij}\, ,
\end{align}
where $i,j=1,2$ and $\vec{x}_1$ and $\vec{x}_2$ satisfy \Eq{eq:lowerbound:d}. The quantity $G(\delta)$, defined by the following expression 
\begin{align}
G(\delta) = \frac{2}{\mathcal{F}^2\left(\delta\right)} \frac{\left(1+\delta\right)^{D+2}-\left(D+2\right)\delta-1}{\left(D+1\right)\left(D+2\right)\delta^2}\, ,
\end{align}
is a prefactor that has been arranged such that, when $\delta\to 0$, $G(\delta)\to 1$. Since the commutator~(\ref{eq:commutnoncan}) is not of the form given by \Eq{eq:commutators:RealSpace}, the fields need to be rescaled according to 
\begin{align}
\label{eq:rescale}
\widetilde{\phi}_{R_i}\left(\vec{x}_i\right) &\equiv \lambda \sqrt{\frac{V_D R_i^D}{G(\delta)}} \phi_{R_i}\left(\vec{x}_i\right)\\
%\qquad \text{and} \qquad
\widetilde{\pi}_{R_i}\left(\vec{x}_i\right) &\equiv \lambda^{-1} \sqrt{\frac{V_D R_i^D}{G(\delta)}} \pi_{R_i}\left(\vec{x}_i\right)\, ,
\end{align}
where $\lambda$ is a prefactor that may be freely set, for instance in order to make $\widetilde{\phi}_R$ and $\widetilde{\pi}_R$ of the same dimension. We will check that our results do not depend on the choice of $\lambda$ anyway, because of local symplectic invariance of the criteria we compute. Then, one can check that 
\begin{align}
\left[\widetilde{\phi}_{R_i}\left(\vec{x}_i\right),\widetilde{\pi}_{R_j}\left(\vec{x}_j\right)\right]=i\delta_{ij}\, ,
\end{align}
hence the rescaled fields are now indeed properly canonically normalised.

The fundamental remark, which is at the basis of the new method presented in this article, is then the following. Let us arrange the rescaled fields, coarse-grained at $\vec{x}_1$ and $\vec{x}_2$, into the four-vector
\begin{align}
\widetilde{\bm Z}_R=\left(\begin{array}{c}
\widetilde{\phi}_{R_1}\left(\vec{x}_1\right)\\ 
\widetilde{\pi}_{R_1}\left(\vec{x}_1\right)\\ 
\widetilde{\phi}_{R_2}\left(\vec{x}_2\right)\\ 
\widetilde{\pi}_{R_2}\left(\vec{x}_2\right)\\ 
\end{array}\right)\, .
\end{align}
In the following, the components of $\widetilde{\bm Z}_R$ will be denoted $\widetilde{Z}_{R,a}$ with $a=1, \cdots , 4$. In a free theory, or in a theory where fields linearly interact with a classical source,  $\phi(\vec{x})$ and $\pi(\vec{x})$ are placed in Gaussian states, and since the coarse-grained fields are constructed as linear combinations of $\phi(\vec{x})$ and $\pi(\vec{x})$, see \Eq{eq:def:CoarseGrain}, they are in a Gaussian state too. As a consequence, $\widetilde{\bm Z}_R$ describes a bi-partite Gaussian system. The first sector, \ie the configuration of the field at $\vec{x}_1$, corresponds to the two first entries of $\widetilde{\bm Z}_R$, while the second sector, \ie the field at $\vec{x}_2$, corresponds to the two last entries of $\widetilde{\bm Z}_R$. One can therefore use the techniques developed for bipartite Gaussian states to characterise the correlations between $\vec{x}_1$ and $\vec{x}_2$. 
\section{Entanglement entropy of Gaussian states}
\label{sec:GaussianStates}
In this section, we recall how the entanglement entropy, the mutual information and the quantum discord of Gaussian systems can be computed. Gaussian states are fully characterised by their two-point correlation function, \ie by the covariance matrix
\begin{align}
\label{eq:defcov}
\gamma_{ab}= 2 \left\langle \left\{ \widetilde{Z}_{R,a}, \widetilde{Z}_{R,b}\right\}\right \rangle \, ,
\end{align}
where $\{,\}$ denotes half the anticommutator, \ie $\{ \widetilde{Z}_{R,a}, \widetilde{Z}_{R,b}\} = (\widetilde{Z}_{R,a} \widetilde{Z}_{R,b} + \widetilde{Z}_{R,b} \widetilde{Z}_{R,a})/2$. If the full quantum system is described by its density matrix ${\rho}_{1,2}$, information about the field configuration at location $\vec{x}_1$ is obtained by tracing over the degrees of freedom corresponding to $\vec{x}_2$, namely
\begin{align}
{\rho}_1 = \mathrm{Tr}_2 \left({\rho}_{1,2}\right)\, ,
\end{align}
and similarly for ${\rho}_2$. The state represented by ${\rho}_1$ is
still Gaussian, and its covariance matrix $\bm{\gamma}_1$ is simply
obtained from $\bm{\gamma}$ by removing the lines and columns
corresponding to $\vec{x}_2$, \ie the third and fourth lines and
columns, so
\begin{align}
\label{eq:gamma12:mat}
\bm{\gamma}_1  =  \left(
\begin{array}{cc}
\gamma_{11} & \gamma_{12}\\
\gamma_{12} & \gamma_{22}
\end{array}
\right) ,
\end{align}
and similarly for $\bm{\gamma}_2$.

Once endowed with the density matrix of a bipartite system, the entanglement entropy, defined as the von-Neumann entropy of either of its subsystems, can be calculated. Concretely, it can be written as
\begin{align}
S_1 = - \mathrm{Tr} \left[{\rho}_1 \log_2\left({\rho}_1\right)\right]\, ,
\end{align}
with similar expressions for $S_2$ and $S_{1,2}$. This general expression is especially easy to evaluate in the particular case of 
a Gaussian state. It is indeed given by~\cite{1999quant.ph.12067H}
\begin{align}
\label{eq:VonNeumann:Entropy:f}
  S({\rho})=\sum_{i=1}^n f(\sigma_i),
\end{align}
where the function $f(x)$ is defined for $x\geq 1$ by
\begin{align}
\label{eq:f(x):def}
  f(x)=\frac{x+1}{2}
  \log_2\left(\frac{x+1}{2}\right)
-\frac{x-1}{2}
\log_2\left(\frac{x-1}{2}\right),
\end{align}
and $\sigma_i$ are the symplectic eigenvalues of the covariance
matrix, that is to say the quantities $\sigma_i$ such that
$\mathrm{Sp}(\bm{J}^{(n)}\bm{\gamma})=\{i\sigma_1,-i\sigma_1,\cdots,
i\sigma_n,-i\sigma_n\}$. In this expression,  $\bm{J}^{(1)}=
  \begin{pmatrix}
    0 & 1 \\
    -1 & 0
  \end{pmatrix}$, and $\bm{J}^{(n)}$ is the $(2n\times 2n)$ block-diagonal
  matrix where each block corresponds to $\bm{J}^{(1)}$, and where $2 n$ is
  the dimension of phase space. From this expression, the mutual information~\cite{6773024}
  \begin{align}
  \mathcal{I} (\vec{x}_1,\vec{x}_2)&= S\left({\rho}_1\right)
  + S\left({\rho}_2\right)
  - S\left({\rho}_{1,2}\right) 
 \label{eq:I:def}
\end{align}
can also be computed, which quantifies the amount of correlations between the field at locations $\vec{x}_1$ and $\vec{x}_2$. 

One way to measure the ``quantumness'' of these correlations is via quantum discord~\cite{2001JPhA...34.6899H, Ollivier:2001fdq}, which corresponds to the difference between two measures of mutual information that coincide for classically-correlated systems, but that may differ otherwise. The first measure of mutual information is the quantity denoted $\mathcal{I}$ and already introduced, while the second measure is
\begin{align}
\label{eq:def:J}
\mathcal{J}(\vec{x}_1,\vec{x}_2)& = S\left({\rho}_1\right)
  - S\left({\rho}_1 \vert \rho_2\right)\, .
\end{align}
In this expression, $S\left({\rho}_1 \vert \rho_2\right)$ is the entropy contained in the first subsystem once the second subsystem has been measured. More precisely, one introduces a complete set of projectors $\{ \Pi_q \}$ along which the second subsystem is measured, where $q$ labels the various projectors. The probability to find the second subsystem in the state on which $\Pi_q$ projects is given by $p_q=\mathrm{Tr}(\rho \Pi_q)$, and through such a measurement the state of the system changes according to $\rho\to \Pi_q \rho \Pi_q/p_q$. The state of the first subsystem after such a measurement is therefore given by $\rho_{1\vert\Pi_q}=\mathrm{Tr}_2 (\Pi_q \rho \Pi_q/p_q)$, which leads to the following expression for the conditional entropy, $S\left({\rho}_1 \vert \rho_2\right) = \sum_q p_q S(\rho_{1\vert\Pi_q})$.  Quantum discord is finally defined as
\begin{align}
\label{eq:discord:def}
  \mathcal{D}(\vec{x}_1,\vec{x}_2)= \min_{\{\hat{\Pi}_q\}}\left[{\cal I}(\vec{x}_1,\vec{x}_2)-{\cal J}(\vec{x}_1,\vec{x}_2)\right]\, ,
\end{align}
where minimisation is performed over all possible complete sets of
projectors, in order to ensure that a non-vanishing discord signals
the presence of genuine quantum correlations for any projection basis.

A generic calculation of quantum discord for Gaussian states is
presented in \Refa{2010PhRvL.105c0501A}. Here we only state
the result in terms of the covariance matrix $\bm{\gamma}$, but a
detailed derivation of the formulas below can be found in that
reference. Let us first denote by
$\bm{\gamma}_{1\mhyphen 2}$ the off-diagonal block of the covariance
matrix, such that the covariance matrix can be written in the block form as
\begin{align}
\bm{\gamma} = \left(
\begin{array}{cc}
\bm{\gamma}_{1} & \bm{\gamma}_{1\mhyphen 2}\\
\bm{\gamma}_{1\mhyphen 2}^\mathrm{T} & \bm{\gamma}_{2}
\end{array}
\right) \, .
\end{align}
The determinant for each block is denoted by $\det\bm{\gamma}_1 \equiv \sigma_1^2$, $\det\bm{\gamma}_2 \equiv \sigma_2^2$ and $\det\bm{\gamma}_{1\mhyphen 2} \equiv \sigma_{1\mhyphen 2}^2$. Quantum discord can be written in terms of these quantities, and
after extremisation over the set of projectors appearing in
\Eq{eq:discord:def}, one has~\cite{2010PhRvL.105c0501A}
\begin{align}
\label{eq:J}
\mathcal{J}(\vec{x}_1,\vec{x}_2) &= f\left(\sigma_1\right)
-f\left(\sqrt{E}\right),
\end{align}
with
\begin{widetext}
\begin{align}
  \label{eq:E}
E = 
\begin{cases}
  \displaystyle
  \frac{1}{(\sigma_2^2-1)^2}\biggl\{2\sigma_{1\mhyphen 2}^4
  +\left(\sigma_2^2-1\right)
    \left(\det{\bm{\gamma}}-\sigma_1^2\right)
    +2\left\vert
    \sigma_{1\mhyphen 2}^2\right\vert \sqrt{\sigma_{1\mhyphen 2}^4
      +\left(\sigma_2^2-1\right)
      \left(\det{\bm{\gamma}}-\sigma_1^2\right)}\biggr\}
\\ \\
\displaystyle
\frac{1}{2\sigma_2^2}\left[\sigma_1^2\sigma_2^2-\sigma_{1\mhyphen 2}^4+\det{\bm{\gamma}}
  -\sqrt{\sigma_{1\mhyphen 2}^8+\left(\sigma_1^2\sigma_2^2-\det{\bm{\gamma}}\right)^2
    -2\sigma_{1\mhyphen 2}^4\left(\sigma_1^2\sigma_2^2+\det{\bm{\gamma}}\right)}\right]\, 
\end{cases}
\end{align}
\end{widetext}
where the first formula applies when $\left(1+\sigma_2^2\right) \sigma_{1\mhyphen
  2}^4\left(\sigma_1^2+\det{\bm{\gamma}}\right)
-\left(\sigma_1^2\sigma_2^2-\det{\bm{\gamma}}\right)^2\geq 0$ and the second
formula otherwise. 

Let us finally mention that, even if the quantum field $\phi(\vec{x})$ is placed in a pure quantum state, because of its \apriori non-trivial real-space correlations, the vector $\widetilde{\bm Z}_R$ generically describes a mixed state. Indeed, when restricting one's
attention to the (coarse-grained) configurations of the field at locations
$\vec{x}_1$ and $\vec{x}_2$, one implicitly traces over its
configuration at all other locations (to which the configurations at
$\vec{x}_1$ and $\vec{x}_2$ are entangled), which leads to a non-pure
bipartite system. In general, this effective ``self-decoherence'' can
be measured with the purity parameter~\cite{PhysRevD.24.1516,
  PhysRevD.26.1862, Joos:1984uk,Colas:2021llj}
\begin{align}
\label{eq:purity:def}
\mathfrak{p} = \mathrm{Tr}(\rho^2)=\frac{1}{\sqrt{\det\bm{\gamma}}}\, ,
\end{align}
which may be used to characterise either the full system
$\rho_{1,2}$ or the reduced systems $\rho_{1}$ and $\rho_2$, by considering
the relevant covariance matrix in each case (namely $\bm{\gamma}$, $\bm{\gamma}_1$ or $\bm{\gamma}_2$). Pure states have
$\mathfrak{p}=1$, while decohered states are such that $0\leq
\mathfrak{p}<1$. 

The above considerations provide all necessary formulas to explicitly compute the entanglement entropy, the mutual information and the quantum discord between the
field configurations at $\vec{x}_1$ and $\vec{x}_2$ from the knowledge of the covariance matrix, which thus achieves our goal.

Before closing this section and illustrating our formalism with a concrete example, let us note that the entries of the covariance matrix, see \Eq{eq:defcov}, can be expressed in terms of the power spectra of the field, which is of practical interest in situations where they can be readily computed. Upon introducing the Fourier transform of the fields,
\begin{align}
\phi(\vec{x})=\left(2\pi\right)^{-D/2}\int\dd^D\vec{k}\, \ee^{-i \vec{k}\cdot\vec{x}} \phi(\vec{k})\, ,
\end{align}
and a similar expression for $\pi(\vec{x})$, one can show that the Fourier moments of the coarse-grained fields defined in \Eq{eq:def:CoarseGrain} are given by $\phi_R(\vec{k}) = \widetilde{W}(kR) \phi(\vec{k})$, where 
\begin{align}
\label{eq:Wtilde:def}
 \widetilde{W}(kR) = \frac{2(D-1)V_{D-1}}{\left(kR\right)^D}\int_0^\infty \dd u\,  W\left(\frac{u}{kR}\right) \sin(u)\,  u^{D-2}\, ,
\end{align}
and the same expression $\pi_R(\vec{k}) = \widetilde{W}(kR) \pi(\vec{k})$ for the conjugate momentum. If the field is placed in a configuration that is statistically homogeneous and isotropic, its two-point function in Fourier space only depends on the modulus of the wave-vector,
\begin{align}
\left\langle \left\{ \phi^\dagger(\vec{k}_1) , \phi(\vec{k}_2) \right\} \right\rangle = \frac{\left(\frac{2\pi}{k}\right)^D}{2(D-1) V_{D-1} } \mathcal{P}_{\phi\phi}\left(k_1\right) \delta\left(\vec{k}_1-\vec{k}_2\right)\, , 
\end{align}
where $\calP_{\phi\phi}$ denotes the reduced power spectrum, and similar expressions define the reduced power spectra $\calP_{\phi\pi}$ and $\calP_{\pi\pi}$. The prefactor in this expression guarantees that 
the two-point correlation function in real space can be 
written as $\langle \{\phi(\vec{x}_1),\phi(\vec{x}_2)\}\rangle 
=\int_0^{\infty} \dd \ln k \, \mathrm{sinc} (k\vert\vec{x}_1-\vec{x}_2\vert){\cal P}_{\phi \phi}(k)$. 

Using \Eqs{eq:rescale} and~(\ref{eq:defcov}), this gives rise to the following formula for the entries of the covariance matrix,
\begin{align}
{\bm{\gamma}}_1 = & 2  \frac{V_D R_1^D}{G(\delta)} \int_0^\infty \dd\ln k \, \widetilde{W}^2(k R_1)
\nonumber \\ & \times
\left(\begin{array}{cc}
\lambda^2  \calP_{\phi\phi}(k) & \calP_{\phi\pi}(k)\\
\calP_{\phi\pi} (k)& \lambda^{-2} \calP_{\pi\pi}(k)
\end{array}
\right), 
\end{align}
with the same expression for $\bm{\gamma}_2$ where $R_1$ is simply replaced by $R_2$, and
\begin{align}
{\bm{\gamma}}_{1\mhyphen 2} = & 
2  \frac{V_D \left(R_1 R_2\right)^{D/2}}{G(\delta)} \int_0^\infty \dd\ln k \, \widetilde{W}(k R_1)\widetilde{W}(k R_2) 
\nonumber \\ & \times
 \mathrm{sinc}\left(k d\right)
\left(\begin{array}{cc}
\lambda^2  \calP_{\phi\phi}(k) & \calP_{\phi\pi}(k)\\
\calP_{\phi\pi} (k)& \lambda^{-2} \calP_{\pi\pi}(k)
\end{array}
\right),
\end{align}
where we recall that $d=\vert \vec{x}_1 - \vec{x}_2 \vert$ denotes the distance between $\vec{x}_1$ and $\vec{x}_2$. The quantities 
$\bm{\gamma}_1$ and $\bm{\gamma}_2$ do not depend on $d$ since they are calculated at the same spatial point, while $\bm{\gamma}_{1\mhyphen 2}$ mixes values of the fields at points $\vec{x}_1$ and $\vec{x}_2$. From these expressions, one can check that neither $\sigma_1$, $\sigma_2$, $\sigma_{1\mhyphen 2}$ nor $\det\bm{\gamma}$ depend on $\lambda$; therefore $\lambda$ cancels out from our final results as announced above. These expressions thus allow one to compute all relevant quantities in terms of the sole power spectra of the field.

\section{Application: massless field in flat space-time}
\label{sec:Minkowski}

In order to illustrate the above formalism with a concrete example, let us consider the case of a massless field in a flat space time. Here, we stress that our main goal is not necessarily to derive original physical results (although the calculation of the quantum discord below is, to the best of our knowledge, new) but to illustrate how our method 
can lead to simple analytical and exact formulas for the relevant quantities while, previously, only numerical techniques would allow 
their determination. For simplicity, we restrict our attention to the case $D=3$, but generalisation to higher dimension is straightforward. The mode functions of the vacuum state are given by $\phi_{\vec{k}} = \ee^{- i k t}/\sqrt{2k}$ and $\pi_{\vec{k}}=\dot{\phi}_{\vec{k}}=-i \sqrt{k/2}\ee^{-i k t}$, which give rise to the reduced power spectra $\calP_{\phi\phi} = k^2/(4\pi^2)$, $\calP_{\pi\pi}=k^4/(4\pi^2)$ and $\calP_{\phi\pi}=0$. The blocks of the covariance matrix can thus be written as
\begin{align}
\label{eq:gamma1:Minkowski}
\bm{\gamma}_1= \frac{2}{3\pi G(\delta)}
\left(
\begin{array}{cc}
\lambda^2 R_1 \mathcal{K}_1 & 0\\
0 & \left(\lambda^2 R_1\right)^{-1} \mathcal{K}_3
\end{array}
\right)
\end{align}
with a similar expression for $\bm{\gamma}_2$ where $R_1$ is simply replaced by $R_2$, and 
\begin{align}
\label{eq:gamma12:Minkowski}
\bm{\gamma}_{1\mhyphen 2}= \frac{2}{3\pi G(\delta)}
\left(
\begin{array}{cc}
 \lambda^2 \sqrt{R_1 R_2} \mathcal{L}_1(\alpha,\kappa) &  0\\
 0 &  \dfrac{\mathcal{L}_3(\alpha,\kappa)}{\lambda^2 \sqrt{R_1 R_2}} 
\end{array}
\right)\, .
\end{align}
In these expressions, we have introduced the two dimensionless parameters
\begin{align}
\alpha\equiv \frac{d}{\sqrt{R_1 R_2}}, \qquad
\kappa \equiv \sqrt{\frac{R_1}{R_2}}\, ,
\end{align}
and the result is expressed in terms of the two integrals
\begin{align}
\mathcal{K}_\mu &= \int_0^\infty z^\mu \, \widetilde{W}^2(z)\, \dd z\, ,\\
\mathcal{L}_\mu\left(\alpha,\kappa\right) &= \int_0^\infty z^\mu \, \widetilde{W}(\kappa z)\, \widetilde{W}\left(\frac{z}{\kappa}\right)\mathrm{sinc}(\alpha z)\, \dd z\, .
\end{align}
Let us note that with the real-space window function introduced in \Eq{eq:WindowFunction:Improved}, the Fourier-space window function $\widetilde{W}$ defined in \Eq{eq:Wtilde:def} is given by 
\begin{align}
\widetilde{W}(z)=\frac{3}{z^3 \mathcal{F}(\delta)} &\left\{\frac{\sin z}{\delta}-\left(1+\frac{1}{\delta}\right) \sin\left[(1+\delta)z\right]
\right. \nonumber \\ & \left.
+\frac{2}{\delta z}\cos(z)-\frac{2}{\delta z}\cos\left[(1+\delta)z\right]\right\}\, .
\end{align}
As a consequence, the integrals $\mathcal{K}_\mu$ and $\mathcal{L}_\mu$ can be expressed in terms of the cosine integral function (we do not give the corresponding expressions since they are not particularly insightful, but they can be readily obtained). 
This allows one to show that, in the limit where $\delta\to 0$, one has 
\begin{align}
\label{eq:integrals:KL}
&  \mathcal{K}_1 = \frac{9}{4}\, ,
\qquad
\mathcal{K}_3 = -\frac{9}{2}\ln \delta\, ,
\\
& {\cal L}_1(\alpha,\kappa)=\frac{3}{160 \alpha \kappa ^5}
    \biggl\{4\kappa^3\left[\alpha ^3 \kappa^2 +11\alpha(1+\kappa^4)\right]
      \nonumber \\ &\quad 
      +{\cal A}_{++}\ln\left(\alpha+\kappa +\frac{1}{\kappa}\right)
    -{\cal A}_{+-}\ln\left(\alpha+\kappa 
    -\frac{1}{\kappa}\right)
        \nonumber \\ &\quad 
    -{\cal A}_{-+}\ln\left(\alpha-\kappa +\frac{1}{\kappa}\right)+{\cal A}_{--}\ln\left(\alpha-\kappa -\frac{1}{\kappa}\right)\biggr\} ,
\\
&\mathcal{L}_3\left(\alpha,\kappa\right)=
\frac{3 }{8 \alpha  \kappa ^3} \left\{4 \, \mathrm{arctanh}\left(\frac{2 \alpha  \kappa ^3}{\alpha ^2 \kappa ^2+\kappa ^4-1}\right)
\right. \nonumber \\ & \quad \left.
+\alpha  \kappa  \left(-2 \alpha ^2 \kappa ^2+6 \kappa ^4+6\right) \mathrm{arctanh}\left(\frac{2 \kappa ^2}{-\alpha ^2 \kappa ^2+\kappa ^4+1}\right)
\right. \nonumber \\ & \quad \left.
+4 \kappa ^6 \, \mathrm{arctanh}\left(\frac{2 \alpha  \kappa }{\alpha ^2 \kappa ^2-\kappa ^4+1}\right)-4 \alpha  \kappa ^3\right\} ,
\end{align}
where the four coefficients ${\cal A}_{++}$, ${\cal A}_{+-}$, ${\cal A}_{-+}$ and ${\cal A}_{--}$ in the definition of ${\cal L}_1(\alpha,\kappa)$ are simple functions of $\alpha$ and $\kappa$ which can be expressed as
\begin{align}
    {\cal A}_{+ +}&=\left[\kappa(\alpha+\kappa)+1\right]^3
    \left[\kappa(\alpha-4\kappa)(\alpha \kappa+\kappa^2-3)-4\right],
    \\
    {\cal A}_{+ -}&=\left[\kappa(\alpha+\kappa)-1\right]^3
    \left[\kappa(\alpha-4\kappa)(\alpha \kappa+\kappa^2+3)-4\right],
    \\
    {\cal A}_{- +}&=\left[\kappa(\alpha-\kappa)+1\right]^3
    \left[\kappa(\alpha+4\kappa)(\alpha \kappa-\kappa^2-3)-4\right],
    \\
    {\cal A}_{- -}&=\left[\kappa(\alpha-\kappa)-1\right]^3
    \left[\kappa(\alpha+4\kappa)(\alpha \kappa-\kappa^2+3)-4\right].
\end{align}
Notice that the condition~\eqref{eq:lowerbound:d} imposes that $\alpha>\kappa+1/\kappa$, which guarantees that the above expressions are always well defined.

From \Eqs{eq:gamma1:Minkowski} and~\eqref{eq:gamma12:Minkowski}, one can see that $\gamma_{12}=\gamma_{34}=\gamma_{14}=\gamma_{23}=0$, $\gamma_{33}=\gamma_{11}/\kappa^2$ and $\gamma_{44}=\kappa^2\gamma_{22}$. This allows us to express all symplectic eigenvalues only in terms of $\gamma_{11}$, $\gamma_{22}$, $\gamma_{13}$ and $\gamma_{24}$. More precisely, one finds that the symplectic eigenvalue of $\bm{\gamma}_1$, which is nothing but $\sigma_1$, and the symplectic eigenvalue of $\bm{\gamma}_2$, which is nothing but $\sigma_2$, are the same, namely
\begin{align}
\sigma_1=\sigma_2=\sqrt{\gamma_{11}\gamma_{22}}=\frac{3}{\pi}\sqrt{-\frac{1}{2}\ln \delta}\, .
\end{align}
On the other hand, the symplectic eigenvalue of the off-diagonal block  $\bm{\gamma}_{1\mhyphen 2}$, namely the quantity $\sigma_{1\mhyphen 2}$, is given by
\begin{align}
\sigma_{1\mhyphen 2}=\sqrt{\gamma_{13}\gamma_{24}}=\frac{2}{3\pi}\sqrt{\mathcal{L}_1\left(\alpha,\kappa\right)\mathcal{L}_3\left(\alpha,\kappa\right)}\, ,
\end{align}
while the full covariance matrix, $\bm{\gamma}$, has two symplectic eigenvalues, given by
\begin{align}
\sigma_\pm = \sqrt{\left(\gamma_{11}\pm\kappa \gamma_{13}\right)\left(\gamma_{22}\pm\frac{\gamma_{24}}{\kappa}\right)}
\underset{\delta\to 0}{\longrightarrow}\sigma_1\sqrt{1\pm\frac{4\mathcal{L}_1\left(\alpha,\kappa\right) }{9}}
\end{align}
in terms of which one has $\det\bm{\gamma}=(\sigma_+\sigma_-)^2$ (this follows from the definition of the symplectic spectrum and the fact that $\det[\bm{J}^{(n)}]=1$). One can check that those symplectic eigenvalues are invariant under the transformation $\kappa\to 1/\kappa$, which is a good sanity check since this simply corresponds to swapping $\vec{x}_1$ and $\vec{x}_2$.

We are now in a position where all relevant quantities can be computed. Let us first focus on one individual susbsystem, \ie the field configuration coarse-grained at a given location. From \Eq{eq:VonNeumann:Entropy:f}, its entanglement entropy is given by $S(\rho_1) = f(\sigma_1)$. Since $f(x)\simeq \log_2(x/2)$ when $x\gg 1$, in the limit $\delta\to 0$ this gives rise to
\begin{align}
S(\rho_1) = \log_2\left(\frac{3}{2\pi}\sqrt{-\frac12 \ln \delta} \right)\, .
\end{align}
Therefore, strictly speaking, the entanglement entropy contained within a hard sphere ($\delta=0$) is infinite for massless scalar fields in the Minkowski background. The purity of the sphere can also be computed, and \Eq{eq:purity:def} gives rise to $\mathfrak{p}_1=\mathfrak{p}_2=1/\sigma_1$, so in the limit $\delta\to 0$ the ``self-decoherence'' effect mentioned above is maximal.
For the mutual information, \Eq{eq:I:def} gives rise to $\mathcal{I}=2f(\sigma_1)-f(\sigma_+)-f(\sigma_-)$, hence
\begin{align}
\label{eq:I:Minkowski}
\mathcal{I}(\vec{x}_1,\vec{x}_2) = -\frac{1}{\ln 4}\ln\left\{1-\left[ \frac{4}{9}\mathcal{L}_1\left(\alpha,\kappa\right)\right]^2\right\}\, ,
\end{align}
where we recall that $\mathcal{L}_1$ is given in \Eq{eq:integrals:KL}. 
\begin{figure}[t]
\begin{center}
\includegraphics[width=0.48\textwidth]{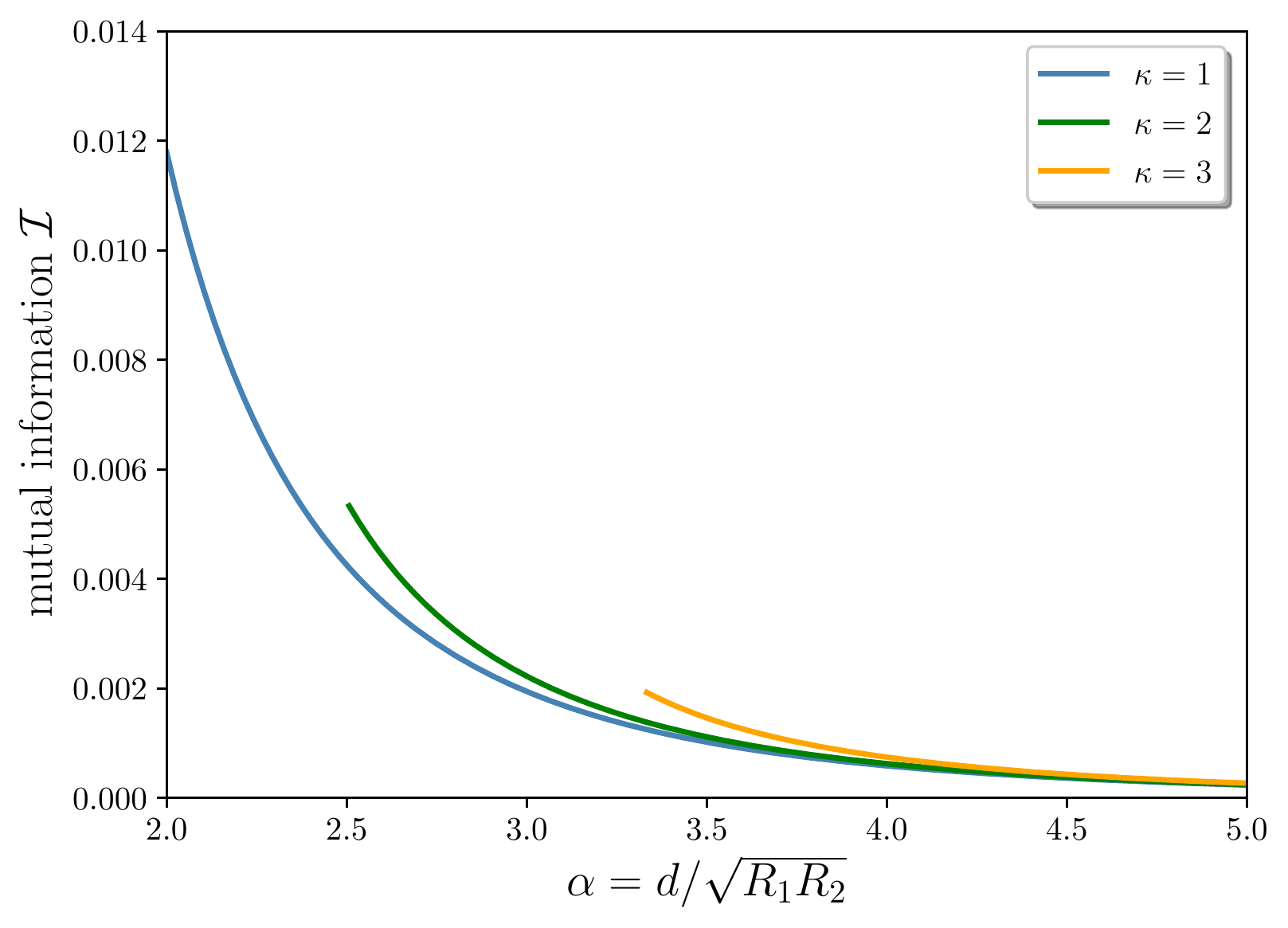}
\caption{Mutual information for a massless scalar field in flat space time, coarse-grained within two spheres of radii $R_1$ and $R_2=R_1/\kappa^2$ and distant by $d$, for a few values of $\kappa$. For the two spheres not to intersect, $\alpha=d/\sqrt{R_1 R_2}$ is restricted to $\alpha>\kappa+1/\kappa$, which is why the curves do not start at the same point.}
\label{fig:I}
\end{center}
\end{figure}
The mutual information is displayed as a function of $\alpha$ in \Fig{fig:I}, for a few values of $\kappa$ (only the case $\kappa\geq 1$ is considered, because of the invariance of the result under $\kappa\to 1/\kappa$). One can see that it decreases with the distance between the two spheres. When $\alpha$ approaches its lowest allowed value, $\alpha\to \kappa+1/\kappa$, the mutual information approaches a finite value that depends on $\kappa$ (we do not give its expression for display convenience, and given that it is easily obtained from the above formulas). In the opposite regime where $\alpha$ is large, one has $\mathcal{L}_1(\alpha,\kappa)\simeq 1/\alpha^2$ and, as a consequence,
\begin{align}
{\cal I}(\vec{x}_1,\vec{x}_2)=\frac{8}{81 \ln 2 }\frac{1}{\alpha ^4}\left(1
+\frac{4}{5 \alpha ^2}\right)+{\cal O}\left(\frac{1}{\alpha^8}\right).
\end{align}
It is interesting to notice that this asymptotic behaviour is independent of $\kappa$, and that one recovers the well-know result~\cite{Shiba:2012np} that the mutual information decays as $(R_1 R_2)^2/d^4$ at leading order. Let us however stress that our formula applies to any distance $d$ between the two spheres and is not restricted to the asymptotic regime. This illustrates well, on this simple and well-known example, the power of the new technique presented here and how it can lead to simple analytical expressions.

Let us now turn our attention to the quantum discord. In the limit $\delta\to 0$, one is in the second case displayed in \Eq{eq:E}, which gives rise to $E=\sigma_+^2\sigma_-^2/\sigma_1^2 \{1+81\sigma_{1\mhyphen 2}^4/[16\sigma_1^4 \mathcal{L}_1^2(\alpha,\kappa)]+\order{\sigma_1^{-8}}\}$, where the expansion is performed at next-to-leading order in $\delta$ (\ie in $1/\sigma_1$) because of a cancellation at leading order for the discord. Indeed, making use of \Eq{eq:J}, one obtains the same expression for $\mathcal{J}$ as the one obtained for $\mathcal{I}$ at leading order, see \Eq{eq:I:Minkowski}. This is why the difference between $\mathcal{I}$ and $\mathcal{J}$, \ie the discord, strictly vanishes in the limit $\delta\to 0$. More precisely, the dominant contribution is of order $\sigma_1^{-2}$, and reads
\begin{figure}[t]
\begin{center}
\includegraphics[width=0.48\textwidth]{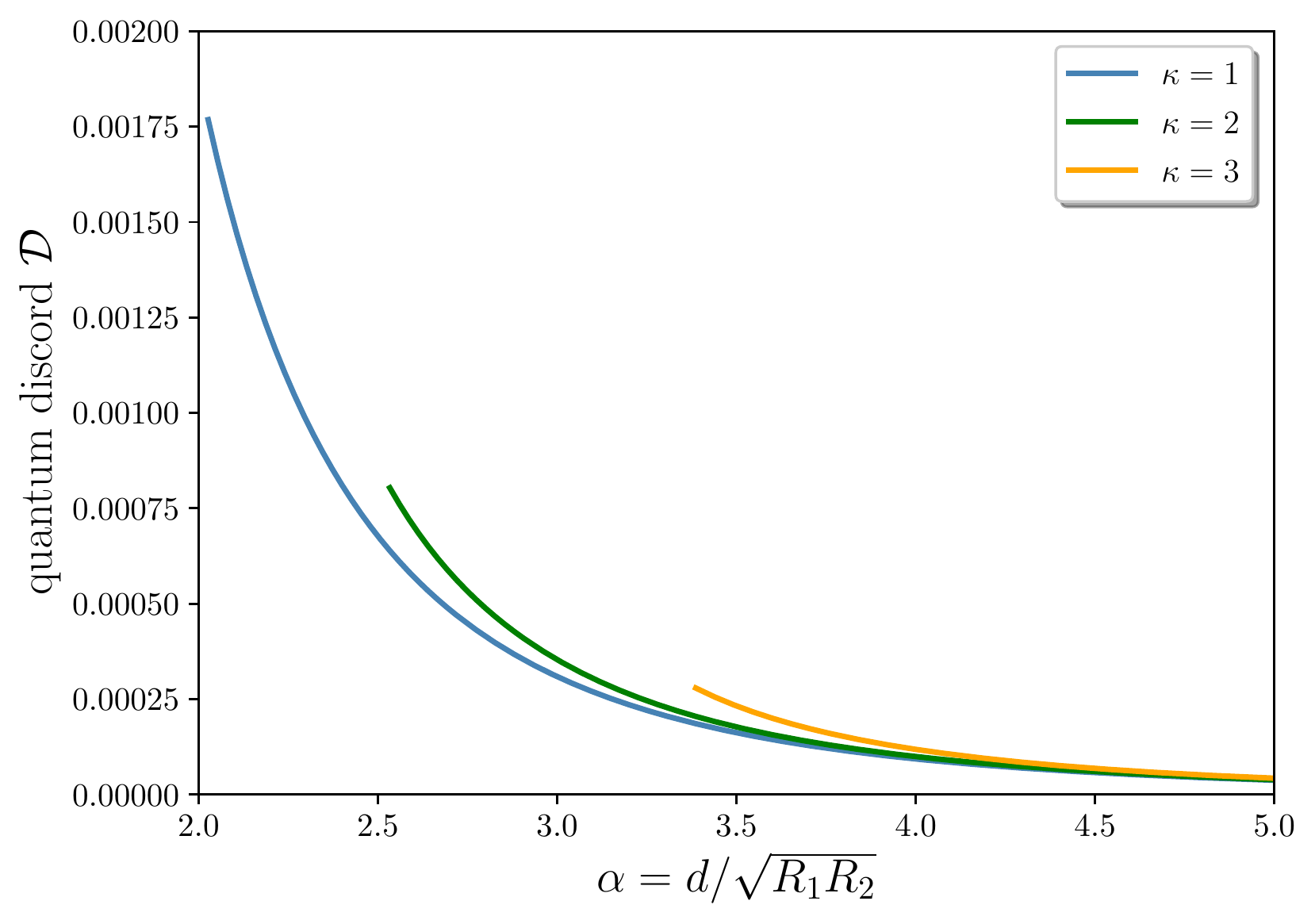}
\caption{Quantum discord for a massless field in flat space time, coarse-grained within two smeared spheres of radii $R_1$ and $R_2=R_1/\kappa^2$ and smearing parameter $\delta=0.01$, and distant by $d$, for a few values of $\kappa$. For the two spheres not to intersect, $\alpha=d/\sqrt{R_1 R_2}$ is restricted to $\alpha>(\kappa+1/\kappa)(1+\delta)$. When $\delta=0$, the discord identically vanishes.}
\label{fig:D:infl}
\end{center}
\end{figure}
\begin{align}
\mathcal{D}(\vec{x}_1,\vec{x}_2)=
\frac{\mathcal{L}_1^2(\alpha,\kappa)}{81/16-\mathcal{L}_1^2(\alpha,\kappa)}\frac{\pi^2}{27\ln 2 \vert\ln(\delta)\vert}\underset{\delta\to 0}{\longrightarrow}0\, .
\end{align}
If $\delta$ is a small, though finite, parameter, the above formula can still be used to evaluate the quantum discord, which is displayed in \Fig{fig:D:infl}. One can see that the behaviour of the discord is similar to the one obtained for the mutual information. In particular, in the large-distance limit, $\alpha\gg 1$, one also obtains $\mathcal{D}\propto \alpha^{-4}$.
\section{Conclusion}
\label{sec:Conclusion}
In this work, we have proposed a new technique to compute the entanglement entropy and all related quantities, such as the mutual information and the quantum discord. This method is applicable to free quantum fields and to fields in interaction with a classical exterior source. Our approach relies on describing the field coarse-grained within several disjoint regions in real space as a multipartite Gaussian system, for which all the relevant tools have been developed over the past few years.
This led us to derive explicit formula that only involve the Fourier-space power spectra of the field.

In order to illustrate how the technique works concretely, we have then applied our formalism to the case of a massless scalar field in flat space-time. We have recovered that the mutual information decays as $(R_1 R_2)^2/d^4$ at large distance $d$ between the two spheres of radii $R_1$ and $R_2$ within which the field is coarse-grained~\cite{Shiba:2012np}, but our formula is in fact exact and, as a consequence, applies even beyond this asymptotic limit, contrary to most calculations of the same kind. We have finally shown that the quantum discord, which measures the amount of ``quantumness'' of the correlations between the two spheres, strictly vanishes if the real-space window function is a top-hat. If the real-space window function is smoother than a top-hat, it decays as $(R_1 R_2)^2/d^4$, like mutual information.

This result that the discord vanishes may be expected from the fact that no particle is created in the vacuum state of the Minkowski background, hence no quantum entanglement builds up. The situation is however more subtle than it seems since the present calculation being performed in real space, one deals with mixed states, for which the connection between entanglement and discord is less straightforward~\cite{2010PhRvL.105b0503G}.

Indeed, let us now consider a more general situation than just a simple free field living in Minkowski spacetime, for instance a scalar field interacting with an homogeneous classical exterior source. In Fourier space, the quantum state of the field, \ie the density matrix, can still be written as a direct product
\begin{align}
\label{eq:rho:separable}
\rho=\underset{\vec{k}\in\mathbb{R}^{3+}}
{\bigotimes}\rho_{\vec{k}}\, .
\end{align}
The fact that $\phi(\vec{x})$ and $\pi(\vec{x})$ are real fields imposes that $\phi(-\vec{k})=\phi^\dagger(\vec{k})$ and a similar relation for $\pi(\vec{k})$ [this explains why the full Hilbert space is labeled by $\vec{k}\in\mathbb{R}^{3+}$, \ie by half of the Fourier space, in \Eq{eq:rho:separable}]. At the quantum level, this implies that particles created with momentum $\vec{k}$ are necessarily entangled with particles with momentum $-\vec{k}$, such as to preserve statistical isotropy. In general, this leads to entanglement between the two sectors $\vec{k}$ and $-\vec{k}$~\cite{Martin:2015qta}, hence to non-vanishing mutual information and to non-vanishing discord. In the Minkowski background, no particle is created, hence both the mutual information and the quantum discord between $\vec{k}$ and $-\vec{k}$ vanish. 
The crucial difference with the real-space calculation is that, here, $\rho_{\vec{k}}$ describes a pure quantum state. This has two main consequences. First, it implies that its von Neumann entropy vanishes, so the mutual information is simply the sum of the von Neumann entropy of the two sub-sectors, and it vanishes when these two sub-sectors are placed in their vacuum state. Second, it also implies that the projected state [denoted $\rho_{1\vert\Pi_q}$ around \Eq{eq:def:J}] is a pure state too (see \eg Appendix A of \Refa{Martin:2015qta}), so its entropy vanishes, and combining \Eqs{eq:I:def}-\eqref{eq:discord:def} leads to $\mathcal{D}=S(\rho_2)=\mathcal{I}/2$ (since $\vec{k}$ and $-\vec{k}$ play a symmetric role). As a consequence, in Fourier space, the vacuum state leads to vanishing mutual information and vanishing discord, while particle creation would produce both mutual information and discord, the former always being twice the later.

In contrast, in real space, as argued around \Eq{eq:purity:def}, the setup consisting of two disjoint compact regions is not placed in a pure state, hence none of these considerations apply. In particular, we find that even in the vacuum state, the mutual information does not vanish (while it does for each Fourier mode individually). It is therefore not \apriori obvious that a vanishing discord in Fourier space translates into a vanishing discord in real space, which is why this result is not a trivial one. Let us also mention that in \Refa{Martin:2021qkg}, we show that this property extends to de-Sitter space-times, where we find that the discord vanishes again if the real-space window function is a top-hap, even though entangled pairs of quanta are created in Fourier space in that case. This illustrates again the non-trivial relationship between entanglement and discord for mixed states~\cite{2010PhRvL.105b0503G}.

Let us finally note that the formalism proposed in this work can be applied to a broader class of situations where Gaussian fields are at play. In the Minkowski background, for instance, one may consider the case of massive fields~\cite{Braunstein:2011sx, Jain:2021ppx} (the only modification of the above calculations would be to replace $k\to \sqrt{k^2+m^2}$), and/or non-scalar fields. One can also consider other types of spatially homogeneous and isotropic backgrounds, such as cosmological backgrounds. In such setups, as already discussed above, particle creation occurs in Fourier space (as an effect of the classical source provided by space-time expansion), and applying our techniques to cosmology is the topic of a separate article, see \Refa{Martin:2021qkg}. Finally, the case of black-holes~\cite{Das:2007mj,Chandran:2020gcd} could be studied with this approach.

\bibliographystyle{JHEP}
\bibliography{Discord_Minkowski}

\end{document}